\documentclass[aps,prl,reprint,groupedaddress,footinbib,nobalancelastpage,showpacs]{revtex4-1}

\usepackage{graphicx}
\usepackage{wasysym}

\begin{document}

\title{Evidence for strong van der Waals-type Rydberg-Rydberg interaction in thermal vapor}

\author{T. Baluktsian}
\thanks{These authors contributed equally to this work.}
\affiliation{5. Physikalisches Institut, Universit\"{a}t Stuttgart,
Pfaffenwaldring 57, 70550 Stuttgart, Germany}
\author{B. Huber}
\thanks{These authors contributed equally to this work.}
\affiliation{5. Physikalisches Institut, Universit\"{a}t Stuttgart,
Pfaffenwaldring 57, 70550 Stuttgart, Germany}
\author{R. L\"ow}
\affiliation{5. Physikalisches Institut, Universit\"{a}t Stuttgart,
Pfaffenwaldring 57, 70550 Stuttgart, Germany}
\author{T. Pfau}
\thanks{t.pfau@physik.uni-stuttgart.de}
\affiliation{5. Physikalisches Institut, Universit\"{a}t Stuttgart,
Pfaffenwaldring 57, 70550 Stuttgart, Germany}

\date{\today}

\begin{abstract}
We present evidence for Rydberg-Rydberg interaction in a gas of rubidium atoms above room temperature.
Rabi oscillations on the nanosecond timescale to different Rydberg states are investigated in a vapor cell experiment.
Analyzing the atomic time evolution and comparing to a dephasing model we find a scaling with the Rydberg quantum number $n$ that is consistent with van der Waals interaction.
Our results show that the interaction strength can be larger than the kinetic energy scale (Doppler width) which is the requirement for realization of thermal quantum devices in the GHz regime.
\end{abstract}


\maketitle

Loosely bound electrons in highlying Rydberg states are giving rise to long range dipolar and van der Waals interactions \cite{Gallagher1988, Saffman2010}.
The coherent control of this interaction allows for engineering of quantum correlated states \cite{Jaksch2000,Lukin2001}.
In ultracold systems where an interaction strength on the order of several MHz is sufficient, the van der Waals interaction has lead to the observation of Rydberg blockade \cite{Tong2004,Heidemann2007} and dephasing \cite{Singer2004,Johnson2008}.
In such systems the interaction has been exploited for quantum logical operations \cite{Isenhower2010, Wilk2010} and the creation of non-classical light states \cite{Dudin2012a}.
So far the observation of this interaction has been limited to the ultracold domain.
In very early related experiments density-dependent line broadening effects on Rydberg lines have been studied \cite{Raimond1981} and interaction effects in thermal vapor involving excited but non Rydberg states have been investigated \cite{Shen2007, Keaveney2012}. 

We present evidence for van der Waals-type interatomic interaction energies in the GHz regime between Rydberg-excited alkali atoms in thermal vapor.
Using a pulsed laser excitation, we are able to drive Rabi oscillations on the nanosecond timescale to a Rydberg state \cite{Huber2011a} and are therefore faster than the coherence time limitation given by the Doppler width.

We investigate the dephasing of these oscillations for different atomic densities and Rydberg states. For a fixed Rydberg state, we see a linear growth of the dephasing rate with density. Through a systematic study of various Rydberg states we have found that the scaling of this growth with the principal quantum number $n$ is consistent with van der Waals-interaction.

\begin{figure}
\centering
\includegraphics[width=\columnwidth]{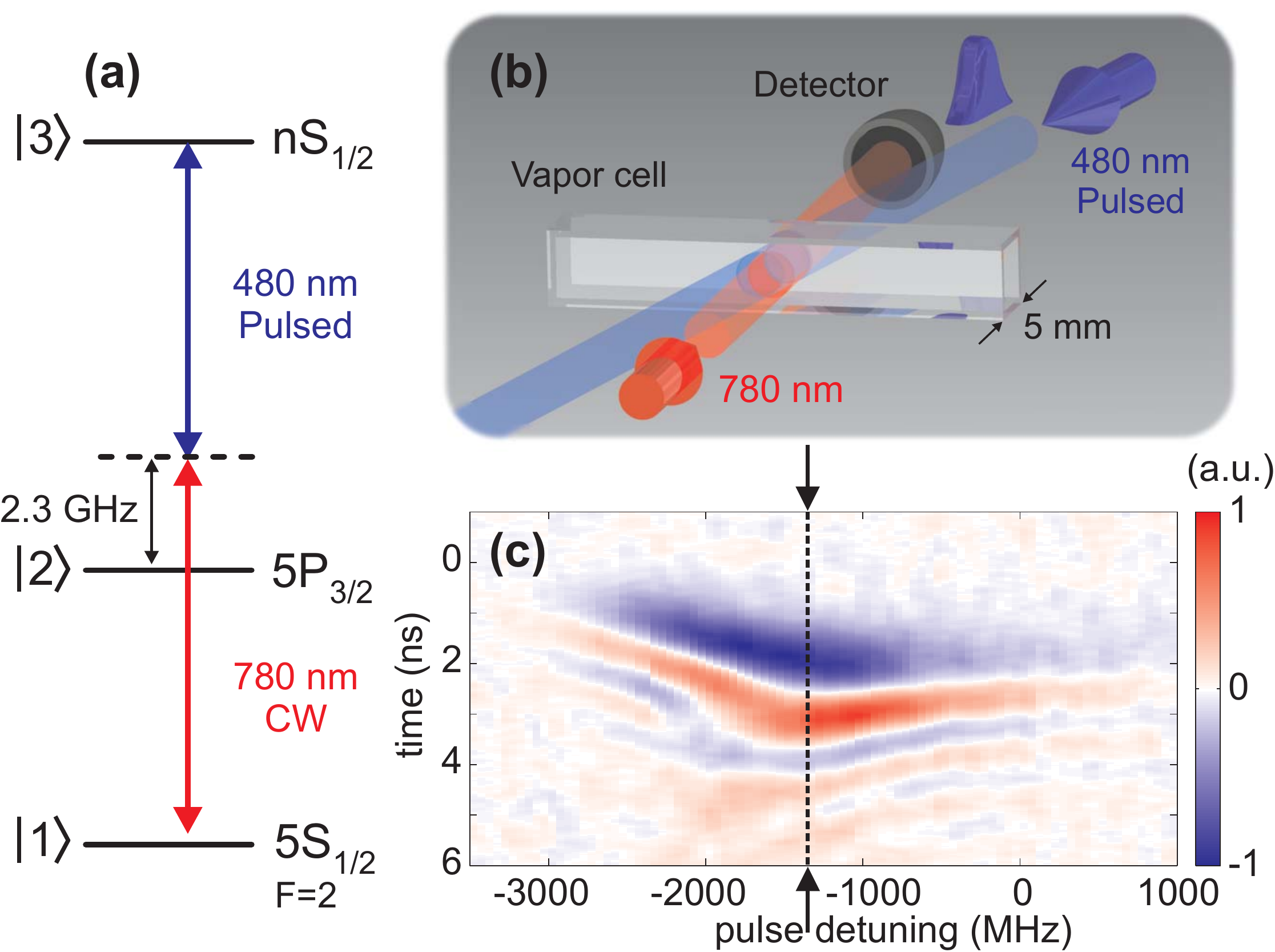}
\caption{Schema of the experiment. 
(a) Relevant level scheme and excitation wavelengths. We use a two-photon-excitation scheme for addressing a Rydberg S-state via the D2-line. The excitation is performed off-resonantly with a (blue) detuning $\Delta/2\pi$ of $\sim 2.3\,\mathrm{GHz}$ with respect to the intermediate state. 
(b) Optical setup. The two beams are overlapped in the vapor cell in almost counter-propagation configuration. We are detecting the transmission of the $780\,\mathrm{nm}$ laser in which the Rabi oscillations manifest themselves during the excitation pulse.
(c) Oscillation signals of a detuning scan of the $480\,\mathrm{nm}$ laser. We fix the frequency position for the analysis to the point of slowest oscillation (dashed line).
\label{fig:schema}}
\end{figure}

We excite $^{85}\mathrm{Rb}$ atoms to a Rydberg S-state with an off-resonant two-photon excitation (Fig.~\ref{fig:schema}a). The upper transition is addressed by a pulsed laser in order to provide sufficient intensity for driving GHz Rabi oscillations despite the small transition dipole matrix element. The peak Rabi frequencies of the pulse on the $5\mathrm{P}-n\mathrm{S}$ transition are in the range of $\Omega_\mathrm{blue}/2\pi \sim 3.3\,\mathrm{GHz}$ to $3.7\,\mathrm{GHz}$. 
The $780\,\mathrm{nm}$ laser addressing the ground state ($\Omega_\mathrm{red}/2\pi\sim 750\,\mathrm{MHz}$) is blue detuned by $\Delta/2\pi \sim  2.3\,\mathrm{GHz}$ with respect to the intermediate state \cite{SuppMat}. 
In this configuration a major part of the population oscillates directly between the ground and Rydberg state. In particular this means that changes in the absorption/emission of the $780\,\mathrm{nm}$ light are mainly caused by these population transfers and not due to excitation of the intermediate state.

A crude estimate of the associated effective two-photon Rabi frequency is given by $\Omega_\mathrm{eff}/2\pi \approx \Omega_\mathrm{blue} \Omega_\mathrm{red}/(2\Delta\cdot 2\pi) = 550\,\mathrm{MHz}$. 
Therefore, the coherent dynamics is faster than the dephasing due to atomic motion expressed by the Doppler width of the two-photon transition of $< 370\mathrm{MHz}$ for our experimental parameters.
Nevertheless it has to be noted that the intermediate level cannot be completely neglected as is done usually in cold atom experiments, as the Rabi frequencies are comparable to the detuning. We will therefore model the experiment using three electronic states.
The exact detuning for the $480\,\mathrm{nm}$ pulse is set to the point of resonant (slowest) Rabi oscillation, which is determined by a detuning scan (Fig.~\ref{fig:schema}c). Note that the two-photon resonance experiences a time-dependent Autler-Townes shift to smaller frequencies due to the strong laser coupling. 

The experimental setup is essentially the same as discussed in \cite{Huber2011a}. The two laser beams are overlapped in a $5\,\mathrm{mm}$ vapor cell in almost counter-propagation configuration (Fig.~\ref{fig:schema}b). The Rabi oscillations during the pulse translate to oscillations in the transmitted intensity of the $780\,\mathrm{nm}$ laser which is subsequently recorded by a fast photodetector. In order to prevent a washing-out of the oscillations due to a distribution of different Rabi frequencies, we use imaging systems before and after the cell to ensure detection from a homogeneously illuminated region.

\begin{figure}
\centering
\includegraphics[width=0.8\columnwidth]{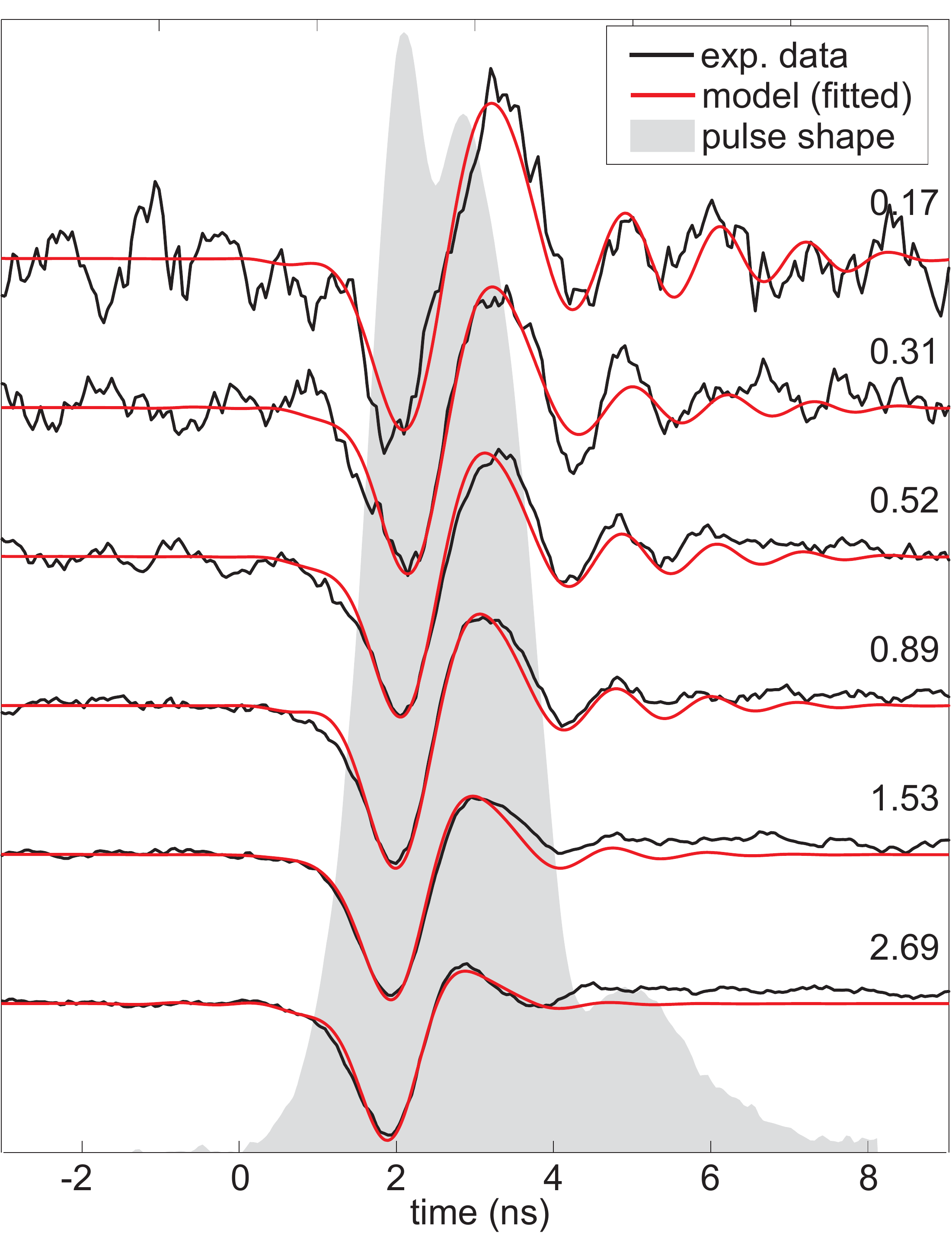}
\caption{Rabi oscillations for different atomic densities. 
Oscillations in the transmission of the $780\,\mathrm{nm}$ laser for different atomic densities are shown here for the Rydberg state 37S. The numbers on the right-hand side give the atomic ground state density in $10^{12}\,\mathrm{cm}^{-3}$. The experimental data (black line) is averaged over 20 shots and normalized in amplitude. A curve derived from a 3-level-atom model (red line) is fitted to the data. The only fit parameters are the dephasing rate of the Rydberg state and the signal amplitude. The gray-shaded curve in the background depicts the measured temporal shape of the excitation pulse intensity.
\label{fig:waterfall}}
\end{figure}

We have been investigating the density dependence of the Rabi oscillations for densities $N_\mathrm{g}$ in the range of $10^{11} - 10^{13}\,\mathrm{cm}^{-3}$ and for different Rydberg S-states. As an example the data for $n=37$ is shown in Fig.~\ref{fig:waterfall}. Note that the curves have been normalized as the original signal amplitude is approximately proportional to the density. We observe an increasing damping in the oscillations with higher density which points towards a dephasing or loss mechanism of the Rydberg state. 

Ionization and Rydberg-ground state collisions do not give a significant contribution on the timescale of our experiment \cite{Thompson1987}. Also incoherent plasma formation in Rydberg gases has been reported to be on a time scale of $\sim 100\,\mathrm{ns}$ or larger \cite{Vitrant1982,Robinson2000} and can be neglected. 
In order to exclude broadening effects on single atoms like e.g. perturbation by electric or electromagnetic fields or collisions with a background gas, we checked for narrow EIT (electromagnetically induced transparency) lines in a CW Rydberg-EIT experiment \cite{Mohapatra2007} with low-power lasers. The observed line widths were less than $20\,\mathrm{MHz}$.
As the Rydberg population in such an EIT experiment is very small (here $< 0.1\%$), Rydberg-Rydberg interactions are not expected to contribute significantly to the broadening of the EIT line \cite{Schempp2010,Pritchard2010}. Having excluded single atom effects, we conclude that the dephasing can only be explained by an interatomic effect that is proportional to the Rydberg population.

In order to quantify the damping behavior we employ a single atom 3-level model for the coherent dynamics \cite{SuppMat}.
Besides the measured pulse shape as an input, the model includes two additional terms that describe a dephasing of the Rydberg state. A constant dephasing term $\Gamma_{\mathrm{offset}}$ accounts for an offset dephasing that occurs in our data. This dephasing varies with the alignment of our pulse amplifier setup which has to be changed for different principle quantum numbers. However for a series of measurements at a particular principle quantum number it is constant. We attribute it - besides atomic motion - to phase noise and imperfections in the spatial homogeneity of the pulse. For the model $\Gamma_{\mathrm{offset}}$ is chosen such that it accounts for all remaining dephasing when extrapolating to zero atomic density. It is on the order of few hundred $\mathrm{MHz}$.
The density-dependent part of the dephasing is described by a term $\rho_{33}\cdot\Gamma_{\rho_{33}}$ that is proportional to the Rydberg population $\rho_{33}$.

\begin{figure}
\centering
\includegraphics[width=\columnwidth]{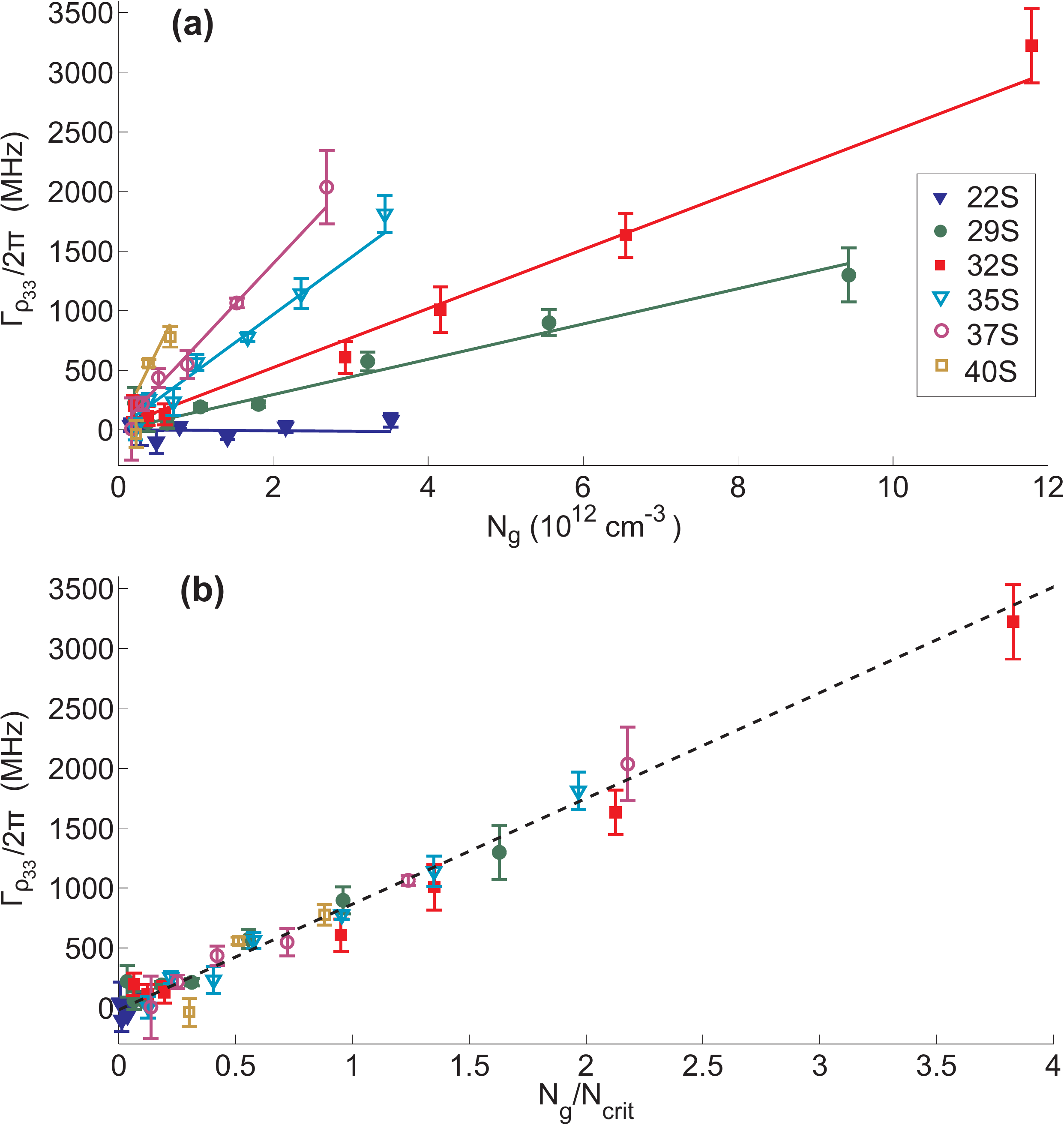}
\caption{Dephasing parameters vs. density. 
(a) The dephasing parameter of the Rydberg state $\Gamma_{\rho_{33}}$ corresponding to the Rydberg-Rydberg interaction is shown as a function of the atomic ground state density $N_{\mathrm g}$. For each Rydberg state the data is fitted by a linear curve.
(b) When scaling the density-axis with the critical density $N_{\mathrm{crit}}$ all data points collapse onto a single line. $N_{\mathrm{crit}}$ is determined by the van der Waals-blockade radius for each Rydberg state individually. The dashed line is a linear fit to all data points. 
\label{fig:dephasings_linear}}
\end{figure}

The dephasing rates $\Gamma_{\rho_{33}}$ are extracted from the experimental data by fitting curves obtained from this model (Fig.~\ref{fig:waterfall}, red lines). The only fit parameters are $\Gamma_{\rho_{33}}$ and the signal amplitude. For each Rydberg state we find a linear increase in the dephasing rate with atomic density $N_\mathrm{g}$ (Fig.~\ref{fig:dephasings_linear}a). The corresponding slope is strongly increasing with the principal quantum number of the Rydberg state. 
Note that the dephasing model only fits well to experimental signals below a certain density (depending on the Rydberg state). For higher densities the model does not accurately describe the data anymore such that it is not possible to extract values for dephasing rates. Whether these effects are due to collective many-body behavior is subject to further study.
The 22S-state shows no increase in the dephasing rate within the margins of error which confirms that our offset dephasing is indeed constant during one series of measurements.

In order to obtain a unified behavior we define a critical density $N_{\mathrm{crit}}$ that corresponds to the interatomic distance at which the van der Waals-shift matches the excitation bandwidth $\Omega_\mathrm{eff}$, which is kept constant for all experiments:
$$\hbar \Omega_\mathrm{eff}\equiv \frac{C_6}{{r_\mathrm{crit}}^6}\equiv C_6 \cdot {N_\mathrm{crit}}^2 .$$
Rescaling the atomic density in units of $N_{\mathrm{crit}}$ for each principal quantum number, we find that all dephasing values collapse onto a single line (Fig.~\ref{fig:dephasings_linear}b).

\begin{figure}
\centering
\includegraphics[width=\columnwidth]{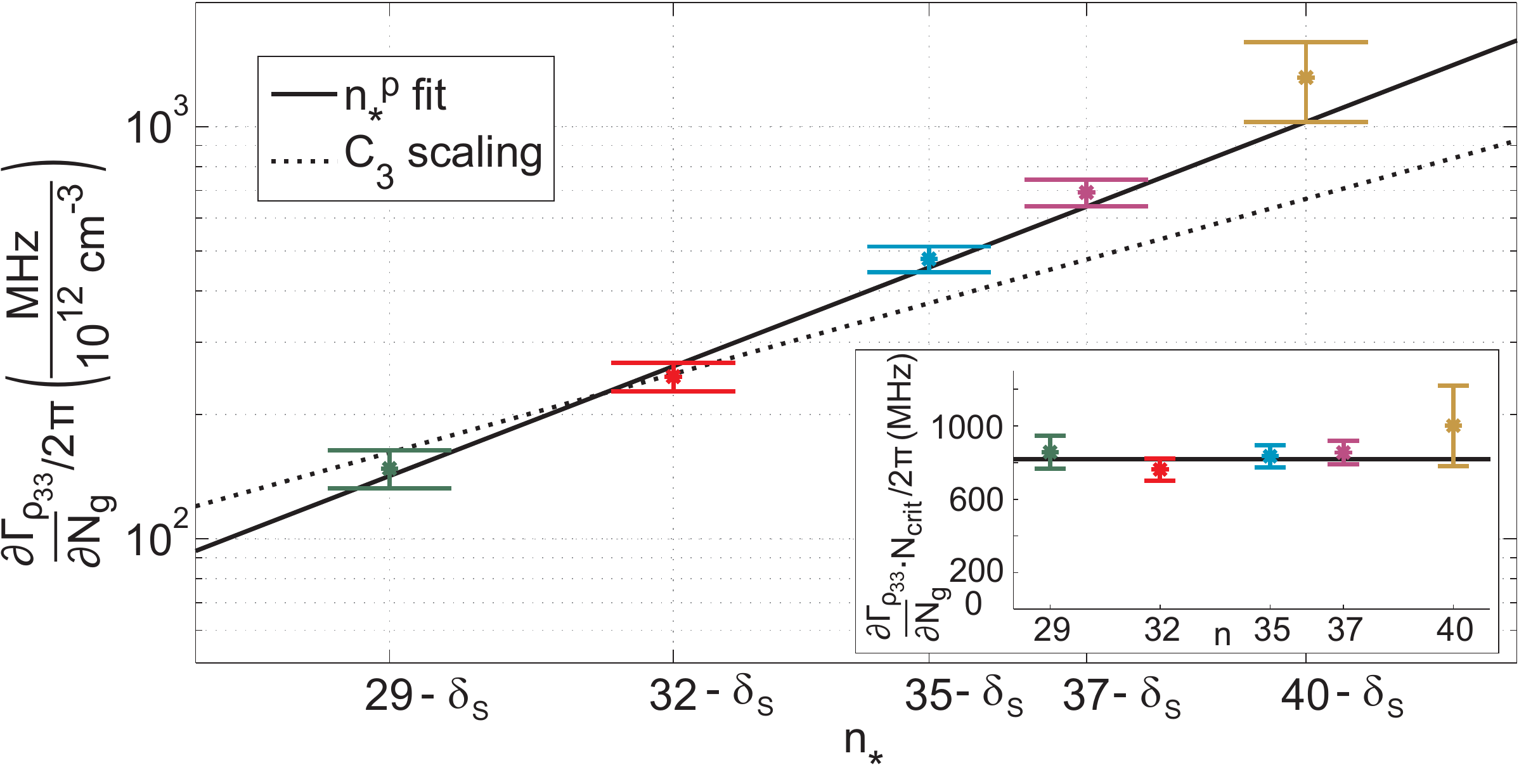}
\caption{Slope of the dephasing vs. principal quantum number. 
Double-logarithmic plot of the dephasing slopes $\partial\Gamma_{\rho_{33}}/\partial N_\mathrm{g}$ vs. the effective principal quantum number $n_*$. 
The fit of a power law $\propto {n_*}^p$ yields an exponent of $p = 5.58 \pm 0.48$ (solid line). This is in excellent agreement with an ${n_*}^{11/2}$ van der Waals scaling. Fitting an ${n_*}^{4}$ scaling curve to the data (dashed line), it can be seen that resonant dipole-dipole interaction is not consistent with the data.
The inset shows the dephasing at the van der Waals critical density. For $N_\mathrm{g} = N_{\mathrm{crit}}$ all examined Rydberg states show the same dephasing.
\label{fig:C6scaling}}
\end{figure}

The van der Waals-shift is expected to behave like
$$\frac{C_6}{r^6}\propto {n_*}^{11}\cdot {N_\mathrm{g}}^2 ,$$
where $n_* = n-\delta_S$ is the effective principal quantum number accounting for the quantum defect $\delta_S \approx 3.13$ \cite{Li2003}.
Accordingly the critical density scales as $N_\mathrm{crit} \propto {n_*}^{-11/2}$.
We therefore would expect the slopes of the dephasing rates to behave like $\partial\Gamma_{\rho_{33}}/\partial N_\mathrm{g} \propto {n_*}^{11/2}$. 
Fitting a power law $\propto {n_*}^p$ to the slopes for the different effective principal quantum numbers (Fig.~\ref{fig:C6scaling}), we obtain an exponent of $p = 5.58 \pm 0.48$ in excellent agreement with van der Waals scaling.

At the interatomic distances of $\sim 1\,\mathrm{\mu m}$ present in the experiment the interaction potentials of different pair states can potentially be within the excitation bandwidth \cite{Cabral2011}. Hence, the perturbative limit which is van der Waals interaction might not necessarily be valid. If a second pair state (e.g. $n'$P$n''$P) with an allowed dipole transition to the respective $n$S$n$S state is within the excitation bandwidth, the interaction energy will also include a $C_3/r^3$ term and thus exhibit ${n_*}^{4}$-scaling \cite{Reinhard2007}. This scaling, however, clearly lies outside the margins of error of the fit and can thus be ruled out. Therefore, we conclude that for the $n$S$n$S states examined the van der Waals description remains valid for interparticle distances down to at least $1\,\mathrm{\mu m}$.

Looking again at the dephasing rates at the van der Waals critical density, we can see that all Rydberg states dephase with the same rate of $\sim 800\,\mathrm{MHz}\cdot \rho_{33}$ (Fig.~\ref{fig:C6scaling}, inset). 
Note that the definition of $N_\mathrm{crit}$ contains a somewhat arbitrary constant factor which has, however, no influence on the $n$-scaling. Therefore, the value at $N_\mathrm{crit}$ is reflecting only the order of magnitude of the interaction energy at the blockaded density.

At higher densities ($N_\mathrm{g} \approx 2 N_\mathrm{crit}$) we have measured dephasing rates of up to $\sim 2\,\mathrm{GHz}$ (for fully excited Rydberg atoms). This value clearly lies outside of the Doppler width of the sample. This shows that the underlying van der Waals interaction strength can be large enough to support a blockade effect in a thermal gas.

In summary we have found evidence of Rydberg-Rydberg interaction in the coherent evolution of atoms in thermal vapor. With a systematic study of the scaling for different Rydberg quantum numbers we could identify the interaction to be of van der Waals type. 
We have seen that the strength of the observed van der Waals interaction can be larger than the Doppler width. This is the basic requirement for the observations of a blockade effect and coherent collective behavior in a thermal ensemble. Hence these results suggest room temperature alkali vapors as promising candidates for quantum devices. 
For the direct observation of a blockade effect further studies in a small spatial volume on the order of the interaction range \cite{Kubler2010} will be necessary.

%Acknowledgements
We acknowledge insightful discussions with J.~Shaffer on the details of the molecular potentials and thank K.~Rz\c a\.zewski and H.~K\"ubler for helpful suggestions. The work is supported by the ERC under contract number 267100, BMBF within QuOReP (Project 01BQ1013), the EU project MALICIA, and contract research 'Internationale Spitzenforschung II' of the Baden-W\"urttemberg Stiftung. T.B. acknowledges support from the Carl Zeiss Foundation. B.H. acknowledges support from Studien\-stiftung des deutschen Volkes.

\newpage

\begin{widetext}
\onecolumngrid
\section{Supplemental Material}
\vspace{0.5cm}
\twocolumngrid
\end{widetext}
%\small

\subsection{Experimental details.}
The ground state laser at $780\,\mathrm{nm}$ is blue detuned by ${\sim 2.3\,\mathrm{GHz}}$ with respect to the center of gravity of the $5\mathrm{P}_{3/2}$ state. This is achieved by locking the laser onto a line of the neighboring $^{87}\mathrm{Rb}$ isotope ($5\mathrm{S}_{1/2}, F=1\rightarrow 5\mathrm{P}_{3/2}$) transition. The $480\,\mathrm{nm}$ laser pulse driving the upper transition is created with a seeded dye amplifier setup in 4-pass configuration similar to the one described in \cite{Schwettmann2007}. The transmitted light at $780\,\mathrm{nm}$ is detected with an AC-coupled avalanche photo diode with a lower cutoff at $\sim 1\,\mathrm{MHz}$. This means only fast changes in the signal are detected. The beam diameters at the position of the cell are $\diameter_\mathrm{blue} = 185$ $\mu$m and $\diameter_\mathrm{red} = 220$ $\mu$m. Further details about the optical setup can be found in previous work \cite{Huber2011a}.

A reservoir containing a droplet of rubidium is attached to the 5 mm vapor cell and can be temperature-controlled. The atomic vapor density inside the cell can be adjusted by changing this temperature. The cell itself is placed in an oven  and kept at a constant temperature of $T \sim 130^\circ C$ in order to prevent condensation of atoms on the glass surface. The exact atomic density in the cell is determined by absorption measurements.
\\

\subsection{3-level model.}
As shown in previous work \cite{Huber2011a}, we model the time evolution of our system with a single atom three-level system with a density matrix approach using the Liouville-von Neumann equation \cite{Fleischhauer2005}
$$\frac{\partial \hat{\rho}}{\partial t} = -\frac{i}{\hbar} \left[ \hat{H}, \hat{\rho}\right] + \hat{L}\left(\hat{\rho}\right)+ \hat{L}_{deph}\left(\hat{\rho}\right).$$
$\hat{H}$ is the Hamiltonian containing the light couplings and $\hat{L}\left(\hat{\rho}\right)$ contains the decays of the levels due to natural linewidths. In order to account for the observed changes in the signal with increasing density we have introduced an additional dephasing of the Rydberg state of the form
\begin{eqnarray*}
\hat{L}_{\mathrm{deph}}\left(\hat{\rho}\right) = \left(\Gamma_{\rho_{33}} \rho_{33}+\Gamma_{\mathrm{offset}}\right)
\left(
\begin{array}{ccc}
0											&0											&-\frac{1}{2}\rho_{13}\\
0											&0											&-\frac{1}{2}\rho_{23}\\
-\frac{1}{2}\rho_{31}	&-\frac{1}{2}\rho_{32}	&0
\end{array}
\right).
\end{eqnarray*}
The dephasing term $\Gamma_{\rho_{33}} \rho_{33}$ is proportional to the Rydberg population and thus describes Rydberg-Rydberg interaction. $\Gamma_{\mathrm{offset}}$ accounts for a constant offset dephasing that we attribute to experimental imperfections. The dephasings only act on the coherences of the density matrix such that no population transfer is induced by them.
For the calculations we have neglected the velocity distribution of the atoms as the dynamics is dominated by the laser detunings ($\Delta_\mathrm{Doppler} \ll \Delta_\mathrm{Laser})$. We have verified that the results with and without Doppler distribution differ by less than $\sim 5\%$.
The time dependence of the blue Rabi frequency $\Omega_\mathrm{blue}$ was extracted from the real temporal envelope of the laser pulse. 
A digital frequency filter was applied to the results to mimic the finite bandwidth ($\sim 1 \mathrm{GHz}$) of our detector.\\

\subsection{Dephasing mechanism.}
\begin{figure}
\centering
\includegraphics[width=\columnwidth]{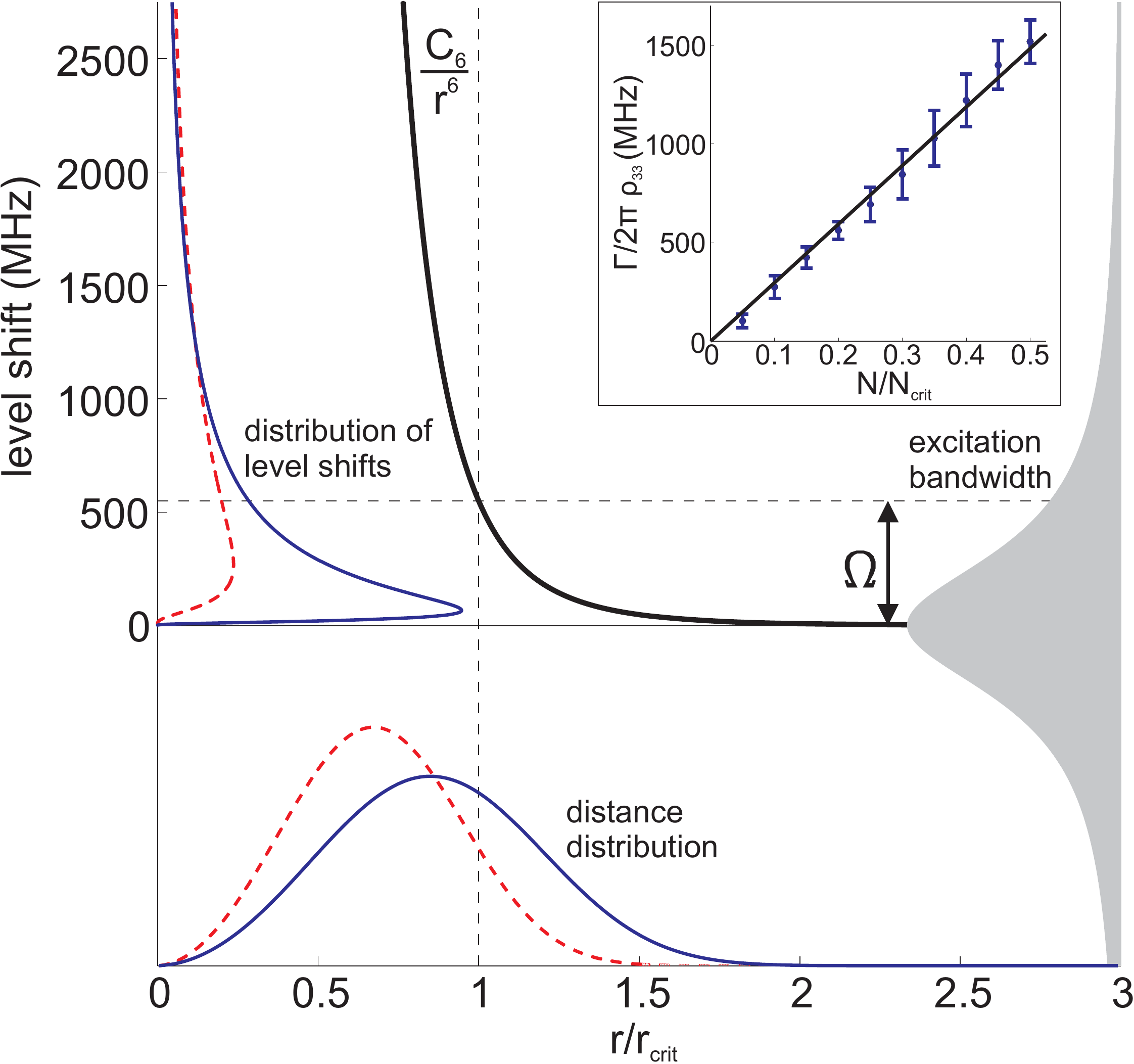}
\caption{Model for the dephasing behavior. 
Each atom experiences a level shift given by the van der Waals interaction potential (black curve). The distance distribution of the atoms is shown in the bottom part for a density of $\mathrm{N}=0.25 \mathrm{N}_\mathrm{crit}$ (blue line) and $\mathrm{N}= 0.5 \mathrm{N}_\mathrm{crit}$ (red dashed line). According to the $1/r^6$-potential this translates to a distribution of Rydberg level shifts shown on the upper left part (blue and dashed red curves, respectively). The excitation bandwidth (gray-shaded curve), which is mainly determined by the effective Rabi frequency $\Omega$, gives a feeling for which level shifts contribute to the signal. The critical density is defined such that the van der Waals shift at the corresponding mean distance $r_\mathrm{crit}$ matches the excitation bandwidth (black dashed lines).
\label{fig:distdist}}
\end{figure}
In order to understand the dephasing mechanism more quantitatively we developed the following simple model. The model is based on an ensemble of atoms, in which the time-evolution of each atom is described by single-atom dynamics only. The Rydberg state of each atom experiences a van der Waals shift given by the distance to its nearest neighbor (Fig.~\ref{fig:distdist}). The distribution of interatomic distances in a thermal vapor is then mapped to a distribution of interaction strengths. Assuming a nearest neighbor distribution of $w(r) = 4\pi r^2 N \exp{\left(-\frac{4\pi}{3}r^3 N\right)}$ for a certain atomic density $N$, we calculate the Rabi oscillations in the signal for each level shift and average then according to the distribution of interaction strengths. The resulting signal is evaluated in the same way as the experimental data by fitting curves from the dephasing model.
The model reproduces the linear density dependence of the dephasing rates for densities $N$ up to $\sim 0.5 N_\mathrm{crit}$ (Fig.~\ref{fig:distdist}, inset). The assumption in the model is that each atom experiences the full shift of the interaction which implies that each nearest neighbor is fully excited to the Rydberg state. 
In the real experiment, however, only an average Rydberg population of $\rho_{33} \sim 20\%-25\%$ (averaged over time) is achieved. This is consistent with the experimental data points ranging until $N_\mathrm{g}\sim 2 N_\mathrm{crit}$. Also the magnitude of the dephasing rate is in good agreement with the experimental results. 

Furthermore it becomes clear that there is a direct connection between the excitation bandwidth and the dephasing rate as only atoms whose shifts lie within the bandwidth can contribute to the signal. The single atom dynamics in the model is justified as long as the fraction of blockaded atoms is small enough as they do not contribute to the signal then. For a significant number of blockaded atoms, however, collective dynamics is expected to appear \cite{Heidemann2007} and the model breaks down.

\end{document}